# Zero-shot learning approach to adaptive Cybersecurity using Explainable AI

Dattaraj J Rao

HEAD AI RESEARCH, PERSISTENT SYSTEMS, LTD. dattaraj_rao@persistent.com

Shraddha Mane

DATA SCIENTIST, PERSISTENT SYSTEMS, LTD. shraddha_mane@persistent.com

Cybersecurity is a domain where there is constant change in patterns of attack, and we need ways to make our Cybersecurity systems more adaptive to handle new attacks and categorize for appropriate action. We present a novel approach to handle the alarm flooding problem faced by Cybersecurity systems like security information and event management (SIEM) and intrusion detection (IDS). We apply a zero-shot learning method to machine learning (ML) by leveraging explanations for predictions of anomalies generated by a ML model. This approach has huge potential to auto detect alarm labels generated in SIEM and associate them with specific attack types. In this approach, without any prior knowledge of attack, we try to identify it, decipher the features that contribute to classification and try to bucketize the attack in a specific category - using explainable AI. Explanations give us measurable factors as to what features influence the prediction of a cyber-attack and to what degree. These explanations generated based on game-theory are used to allocate credit to specific features based on their influence on a specific prediction. Using this allocation of credit, we propose a novel zero-shot approach to categorize novel attacks into specific new classes based on feature influence. The resulting system demonstrated will get good at separating attack traffic from normal flow and auto-generate a label for attacks based on features that contribute to the attack. These auto-generated labels can be presented to SIEM analyst and are intuitive enough to figure out the nature of attack. We apply this approach to a network flow dataset and demonstrate results for specific attack types like ip sweep, denial of service, remote to local, etc.

CCS CONCEPTS • Software security engineering • Network security • Artificial intelligence

**Additional Keywords and Phrases:** Explainable AI, Shapely explanations, alarm-flooding

## 1 INTRODUCTION

Modern security information and event management (SIEM) and intrusion detection (IDS) systems leverage Machine Learning (ML) to correlate network features, identify patterns in data and highlight anomalies corresponding to attacks. Security researchers spend many hours understanding these at-tacks and trying to classify them into known kinds like port sweep, password guess, teardrop, etc. However, due to the constantly changing attack landscape and emergence of advanced persistent threats (APT), hackers are continuously finding new ways to attack systems. A static list of classification of attacks will not be able to adapt to new and novel tactics adopted by adversaries. Also due to the constant flow of alarms generated by multiple sources in the network it be-comes difficult to distinguish and prioritize particular types of attacks. This is the classic alarm flooding problem. A possible solution to this would be if we had a smart system that could auto-label alarms and categorize them so the SIEM analyst can focus on particular alarm types. To try and resolve this, we propose a dynamic classification system using zero-shot classification approach to ML.

Zero-shot learning, and one-shot learning are emerging technique in Computer Vision (CV) and Natural Language Processing (NLP) to model problems where novel classes may be identified during runtime rather than models being explicitly trained to recognize them. In CV, one-shot learning is often used to save facial encodings of people in a database and then use the same neural network to create encoding on new face and identify face matching closest to a stored encoding [11][12][13]. This technique also has been used to match other visual elements like railway tracks and switches [14]. In NLP, since the feature space and class space are in same do-main, we could jointly relate the class names to feature sentences to directly do classification in a zero-shot setting [15][16][17].

In this paper, we take a novel approach to merge zero-shot learning with explainability study of ML models. ML systems often tend to be black boxes and do not do a very good job giving some evidence as to why they made a particular classification or value prediction. Recent research in understanding the reasoning behind predictions of ML models has led to rise of a new field called explainable AI (XAI). Organizations like DARPA [18] have invested heavily in this study to make ML models more ethical and unbiased. Several approaches and libraries have been published that can analyze black-box mod-els and generate explanations [19]. We will particularly focus on a library called SHAP that is inspired by game-theory to generate explanations [10].

Let's see an example of classification of an attack using IDS data. For example, if we classify an attack as a 'guess pass-word' and provide evidence that this is because the number of hot indicators is 1 and the source bytes is around 125 – it gives much better visibility to the security analyst on why the alert is flagged. Also, if the explanation matches the domain knowledge the analyst can easily approve it with confidence. Moreover, for a new attack when the anomaly is flagged and explanation given, the analyst may decide if this is a new pat-tern of unknown attack and possibly capture this as a rule in a system like Zeek (formerly Bro-IDS) [20].

This analysis may eventually be combined with a natural language generating system that can provide concrete English statements explaining the alert and reasons for it. Figure 1 below shows the high-level schematic of how the system will work and key components. We will use a zero-shot learner to predict labels of unknown classes after they are predicted as an attack using our pre-trained ML model. Then we show these auto-generated labels to security analysts for verification. The label should give enough information for security experts to understand some details of the attack. They may further assign a better label to this attack which will be stored and used henceforth as a new label for that attack type.

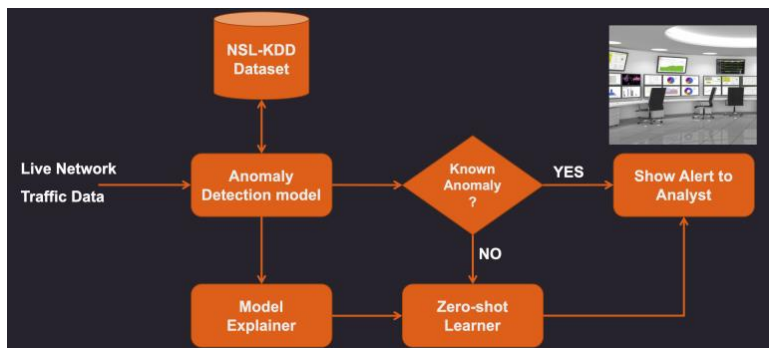

**Fig. 1.** Schematic of the system

## 2 ANALYSIS OF THE DATA

We will use the widely popular NSL-KDD dataset [1] which contains several popular network intrusion attacks with key features captured using a network flow analysis tool and then labelled manually. Below figure 2 shows comparison of Training (Data frame) and Test (Compared) datasets volume in the NSL-KDD dataset. This analysis is done using the SweetViz tool [9].

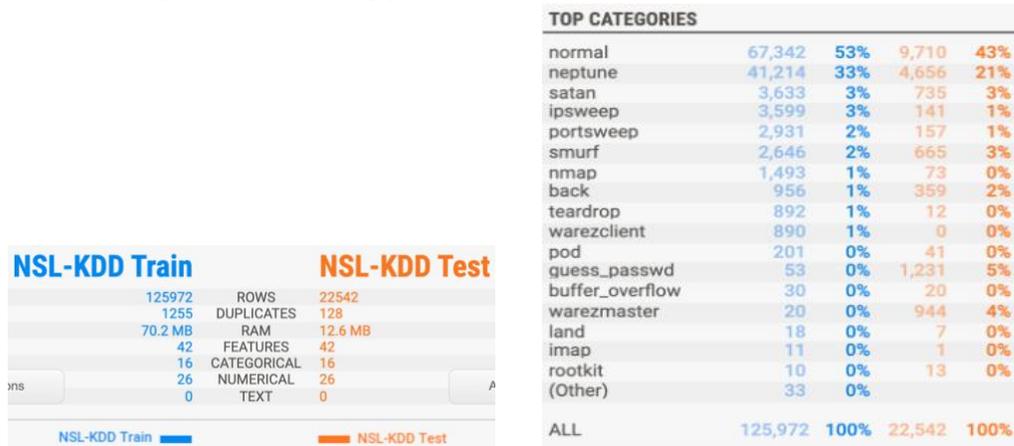

Fig. 2. Basic overview of NSL-KDD dataset

The type of attacks in NSL-KDD dataset are also shown in figure 2. NSL-KDD data set covers four major categories of attacks such as Probing attacks (information gathering attacks), Denial-of-Service (DoS), User-to-Root (U2R) attacks (unauthorized access to local super-user or root), and Remote-to-Local (R2L) attacks (unauthorized local access remotely). The actual labels in the dataset show sub-categories of the attacks like portsweep, neptune, satan, ipsweep, etc. Each attack has a significant pattern defined by the features in the dataset which our ML model tries to learn. After we train the model with acceptable accuracy (above 90%), we will decode these feature patterns learned buy the model using explainable AI to understand type of attack. This way, we try and establish a generalized scheme for labelling even future unknown attacks.

## 3 CLASSIFICATION ML MODEL

We will classify any data point from NSL-KDD that is not labelled as "normal" as an anomaly. We want to train a generic classification model and use that for generating explanations. Our objective will be to first predict that the data point is actually an attack – then using explanations generated for that attack, classify the attack directly in the auto-generated class name. The class name label will be auto generated using a combination of features that are important in attack prediction and using this combination to generate the attack label.

This is important because for newer attacks we could directly generate the classify the attack type without prior training (zero-shot). The class names which include the key features contributing to attack should be intuitive for an analyst to understand the type of attack. Then we could map the auto generated attack names to the actual attack names by taking feedback from the security analyst. We trained an isolation forest anomaly detection algorithm on the NSL-KDD dataset. This algorithm 'isolates' observations by recursively splitting the data into parts based on a random threshold value till points are isolated. Anomalies are points with shortest splits since they are different from rest. We see that the model gets an acceptable f1 score of 0.92 and 0.91 on normal and attack labels as shown in the table below.

```
              precision    recall  f1-score   support

           0       0.91      0.93      0.92      6776
           1       0.92      0.89      0.91      5822

    accuracy                           0.92     12598
   macro avg       0.92      0.91      0.91     12598
weighted avg       0.92      0.92      0.92     12598
```

**Fig. 3**. Classification report for Isolation Forest on NSL-KDD training data

## 4  GENERATING EXPLANATIONS

SHAP [10] summary plot is global explanation of a model which combines feature importance with feature effect. Shapely value for a feature and particular sample is represented by a point on a summary plot. Features are on Y axis and shapely values are on X-axis. Colors are used to represent low/high values. Features are arranged according to their importance, top feature in the summary plot is most important whereas bottom one is the least.

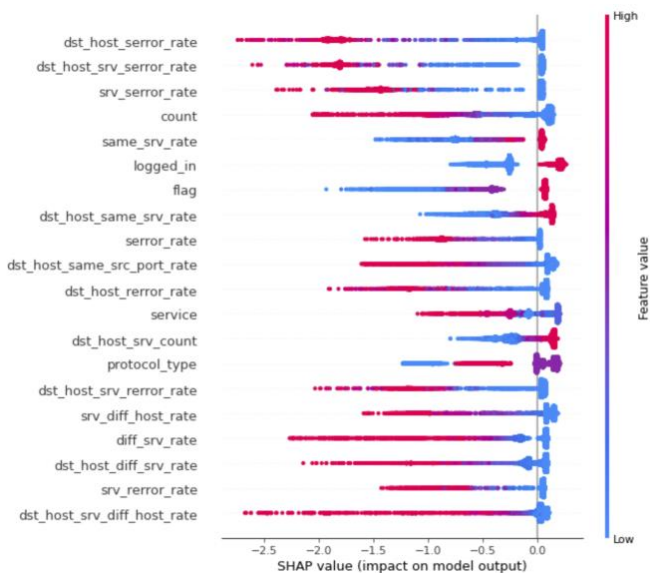

**Fig. 1.** SHAP summary plot for isolation forest model

SHAP global explanations are drawn by considering complete/partial dataset. SHAP local explanations consider only specific instance at a time and generated explanation, it shows which feature values are taking decision towards positive and which are taking towards negative. Figure 6 shows local explanation where probability of the output being attack is 1.00 and features along with their values are shown below such as 'dst_host_same_srv_rate', 'same_srv_rate', 'service_private' and so on. Features pushing the prediction higher are shown in red and those pushing predictions to lower are shown in blue. Note that these explanation changes as we change the input instance.

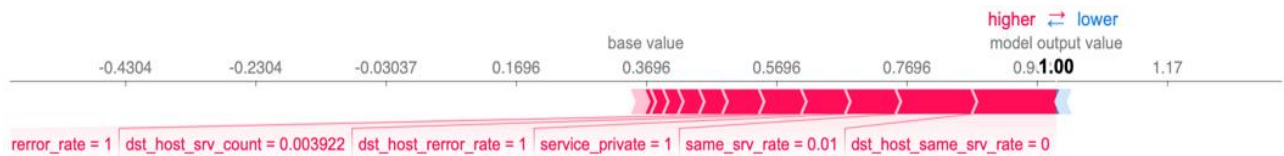

**Fig. 2.** SHAP force plot used for local explanations to explain a particular instance where output probability of 'attack' is 1 and shows features contributing in decision

We have trained a ML model to identify if the network traffic is 'normal' or 'attack'. We see from SHAP values that distinct attack types show different dependencies on features while predicting attack. We will use a BRCG [8] algorithm that can summarize the model prediction by using rules. We use BRCG to extract rules out of the data as follows:

**Predict Y=1(Attack) if ANY of the following rules are satisfied, otherwise Y=0 (Normal):**

· wrong_fragment > 0.00

· src_bytes <= 0.00 AND dst_host_diff_srv_rate > 0.01

· dst_host_count <= 0.04 AND protocol_type_icmp

· num_compromised > 0.00 AND dst_host_same_srv_rate > 0.98

· srv_count > 0.00 AND protocol_type_icmp AND service_urp_i not

Performance of the BRCG algorithm:

**Training Accuracy = 0.9823**

**Test Accuracy = 0.7950**

Which means by applying just these rules one can get ~80% accurate results on unseen test data. We can generate these rules with predictions and have these verified with security domain experts to make sure our model has captured the domain correctly. However, to get to a zero-shot capability we need to generate labels for detected anomalies using explanations. For this we will explore local explanations that try and explain individual predictions. We will use a library called LIME (Local Interpretable Model-Agnostic Explanations) [7] for this.

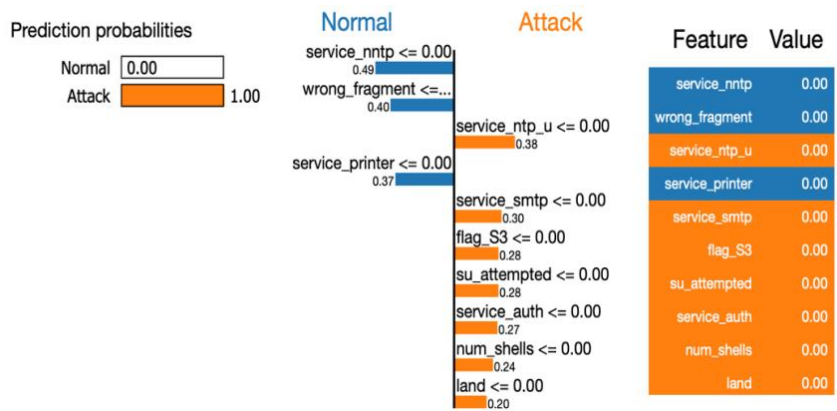

**Fig. 3.** Explaining individual prediction of deep learning classifier using LIME [7]

LIME [7] generates local explanations. In the following figure explanation is shown to determine if classification result is 'Normal' or 'Attack' along with probability and original instance values. Colors are used to highlight which features contributes to which class. Features in orange colour contributes to 'attack' and blue contributes to 'normal' category.

|  | 0 | 1 | 2 | 3 | 4 |  | 0 | 1 | 2 | 3 | 4 |
|---|---|---|---|---|---|---|---|---|---|---|---|
| duration | 0 | 0 | 0 | 0 | 0 | duration | 1.0 | 1.00 | 1.0 | 1.0 | 1.0 |
| src_bytes | 0 | 7.47846e-07 | 0 | 0 | 0 | src_bytes | 1.0 | 0.08 | 1.0 | 1.0 | 1.0 |
| dst_bytes | 0 | 0 | 0 | 0 | 0 | dst_bytes | 1.0 | 1.00 | 1.0 | 1.0 | 1.0 |
| land | 0 | 0 | 0 | 0 | 0 | land | 1.0 | 1.00 | 1.0 | 1.0 | 1.0 |
| wrong_fragment | 0 | 0 | 0 | 0 | 0 | wrong_fragment | 1.0 | 1.00 | 1.0 | 1.0 | 1.0 |
| ... | ... | ... | ... | ... | ... | ... | ... | ... | ... | ... | ... |
| flag_S3 | 0 | 0 | 0 | 0 | 0 | flag_S1 | 1.0 | 1.00 | 1.0 | 1.0 | 1.0 |
| flag_SF | 0 | 1 | 0 | 0 | 0 | flag_S2 | 1.0 | 1.00 | 1.0 | 1.0 | 1.0 |
| flag_SH | 0 | 0 | 0 | 0 | 0 | flag_S3 | 1.0 | 1.00 | 1.0 | 1.0 | 1.0 |
| Class | Attack | Attack | Attack | Attack | Attack | flag_SF | 1.0 | 0.08 | 1.0 | 1.0 | 1.0 |
| Weight | 0.935025 | 3.00021e-05 | 0.000150011 | 0.0255018 | 0.0392928 | flag_SH | 1.0 | 1.00 | 1.0 | 1.0 | 1.0 |

**Table 1a.** Similar instances predicted as attack **1b.** Use of weights to show similarity

The person who takes final decision based on model's output can get understanding of model's decision if we show instances from the training dataset which are similar in different ways to test instance we want to understand. We considered first instance from test dataset for which model prediction is 'attack'. Table 3a shows similar instances from training data, similarity is indicated by the weight mentioned in last row.

It also provides human friendly explanations showing feature values in terms of weight. More the weight, more the similarity. Above two tables- table 3a and table 3b represent five closest instances to the test instance. Based on the weights mentioned, we can see that instance under column 0 is the most representative of test instance as weight it 0.93. These tables would help the analyst to take final decision more confidently.

For end users, ML models should be transparent. They should get answers to their all queries such as why model made certain decision, which factors led to this decision, by making what changes model's decision can be changed etc. CEM algorithm helps us to answer all these end user questions.

We considered one particular instance where prediction made was 'normal', CEM shows us how decision can be changed by making minimal changes in the feature values.

CEM can also highlight minimal set of features along with their values that would maintain the prediction made by the model. Looking at the statistics of explanations given by the CEM algorithm over bunch of applicants, one can get insight into what minimal set of features play important role. It is also possible to get values of these features for every type of attack.

Sample: 2
prediction(X) [[1.00e+00 3.15e-10]] Normal
prediction(Xpn) [[0.18 0.82]] Attack

|  | X | X_PN | (X_PN - X) |
|---|---|---|---|
| duration | 3.4653e-05 | 0.02 | 0.02 |
| hot | 0 | 0.12 | 0.12 |
| dst_host_serror_rate | 0 | 0.03 | 0.03 |
| Class | Normal | Attack | NIL |

PP for Sample: 5
Prediction(Xpp) : Normal
Prediction probabilities for Xpp: [[0.55 0.45]]

|  | X | X_PP |
|---|---|---|
| num_root | 0 | 0.02 |
| count | 0.00782779 | 0.03 |
| srv_count | 0.00782779 | 0.06 |
| diff_srv_rate | 0 | 0.02 |
| dst_host_count | 0.607843 | 0.01 |
| Class | Normal | Normal |

Table 2. Pertinent negative and pertinent positives for an instance

## 5. CONCLUSIONS

Figure 6 below shows the SHAP force plot for group of points from test dataset. We combined 50 points from each category - normal and 3 types of attacks and plotted a force plot for this shown below. This is basically created by taking multiple force plots for a single instance (shown in figure 5) rotating them by 90 degrees and stacking them horizontally. We see a clear separation on type of attack defined by the explanations. We see that between the isolation forest anomaly detector is able to separate between normal and anomalous points. Then using SHAP explainer we can auto-assign labels to the predicted labels using combination of features that appear as important. The SHAP scores show a distinct pattern of explanations for each type of anomaly.

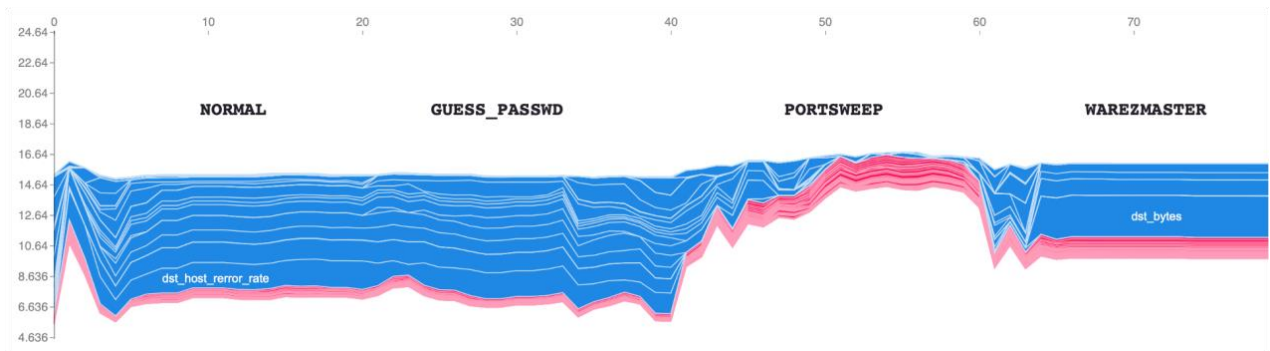

**Fig. 4.** SHAP force plot for 4 types of data points in NSL-KDD dataset

We combined data in NSL-KDD for normal traffic flow and 3 types of attacks and plotted a force plot for this shown below is figure 6. We see a clear separation on type of attack defined by the explanations. Prominence of blue values indicate tendency towards prediction of 0 indicating a normal traffic flow. When red values are prominent it indicates an attack and using explainability we see the features that are more influencing the value being red or 1 – indicating attack. The red patterns for attacks also show a clear separation based on type of attacks. This gives us confidence that the labels generated will have enough uniqueness to satisfy a zero-shot learning requirement. Now we can auto-generate labels indicating a type of attack. Table 1 below shows the auto-generated labels corresponding to above attack types. We selected 5 random points for each attack and show the actual and auto-generated label.

| ACTUAL LABEL | AUTO-GENERATED LABEL |
| --- | --- |
| guess_passwd | dst_host_rerror_rate-hot-service |
| guess_passwd | dst_host_rerror_rate-hot-service |
| guess_passwd | dst_host_rerror_rate-hot-service |
| guess_passwd | dst_host_rerror_rate-hot-service |
| guess_passwd | dst_host_rerror_rate-hot-service |
| portsweep | dst_host_same_srv_rate-service-src_bytes |
| portsweep | dst_host_same_srv_rate-service-src_bytes |
| portsweep | dst_host_same_src_port_rate-service-src_bytes |
| portsweep | dst_host_same_srv_rate-service-src_bytes |
| portsweep | dst_host_same_srv_rate-service-src_bytes |
| portsweep | dst_host_same_srv_rate-service-src_bytes |
| warezmaster | dst_host_count-dst_host_rerror_rate-src_bytes |
| warezmaster | dst_host_count-dst_host_rerror_rate-src_bytes |
| warezmaster | dst_host_count-dst_host_rerror_rate-src_bytes |
| warezmaster | dst_host_count-dst_host_rerror_rate-src_bytes |
| warezmaster | dst_host_count-dst_host_rerror_rate-src_bytes |

**Table 1.** Mapping the actual and auto-generated labels for few random data points

We see from above that the auto-generated label is uniquely mapped to actual label – except for a case for portsweep. Even though there are multiple auto-generated labels for a unique attack, we could allow multiple mappings to the same attack in our learning system. For the portsweep example above both the generated labels – "dst_host_same_srv_rate-service-src_bytes" and "dst_host_same_src_port_rate-service-src_bytes" have a domain meaning which a security analyst can understand. This is particularly helpful for new and unseen attacks since we could auto generate labels like this in a zero-shot learning setting and make our cybersecurity systems more adaptive to novel attack types.

NSL-KDD is a pretty old and exhausted dataset in cybersecurity with all the attacks known and labelled. We are working with a major cybersecurity provider to build this auto-labelling system on a real-world product which sees unknown attacks and is used in a dedicated security operations center (SoC). This will help us evaluate the true potential of this technology and how it can help improve the cyber defense capability.

cing